\documentclass[12pt]{article}
\usepackage{epsf,amsfonts,amsthm,amsmath}

\epsfverbosetrue
\makeatletter
\def\@citex[#1]#2{\if@filesw\immediate\write\@auxout
        {\string\citation{#2}}\fi
\def\@citea{}\@cite{\@for\@citeb:=#2\do
        {\@citea\def\@citea{,}\@ifundefined
        {b@\@citeb}{{\bf ?}\@warning
        {Citation `\@citeb' on page \thepage \space undefined}}
        {\csname b@\@citeb\endcsname}}}{#1}}
\newif\if@cghi
\def\cite{\@cghitrue\@ifnextchar [{\@tempswatrue
        \@citex}{\@tempswafalse\@citex[]}}
\def\citelow{\@cghifalse\@ifnextchar [{\@tempswatrue
        \@citex}{\@tempswafalse\@citex[]}}
\def\@cite#1#2{{\if@cghi\unskip$\null^{#1}$\else #1\fi\if@tempswa\typeout
        {warning: optional citation argument ignored: `#2'} \fi}}

\def\@biblabel#1{$\null^{#1}$}

\makeatother

\usepackage{fontenc,indentfirst, delarray,amsfonts,amsmath,amssymb}
\usepackage{graphicx}
\DeclareGraphicsExtensions{ps,eps,ps.gz}

\begin{document}

\newtheorem{lem}{Lemma}
\newtheorem{thm}[lem]{Theorem}
\newtheorem{prop}[lem]{Proposition}
\newtheorem{cor}[lem]{Corollary}

\newcommand{\norm}[1]{\left\lVert#1\right\rVert}
\newcommand{\abs}[1]{\lvert#1\rvert}
\newcommand{\scal}[1]{\langle#1,#1\rangle}
\newcommand{\scl}[2]{\langle#1,#2\rangle}
\newcommand{\suup}[1]{ \underset{#1}{\sup} }
\newcommand{\grad}[1]{\overrightarrow{\triangledown} #1}
\newcommand{\ket}[1]{\lvert#1\rangle}

\font\twelve=cmbx10 at 13pt
\font\eightrm=cmr8
\baselineskip 18pt

\def\kk{{\mathbb{K}}}
\def\cc{{\mathbb{C}}}
\def\rr{{\mathbb{R}}}
\def\nn{{\mathbb{N}}}
\def\zz{{\mathbb{Z}}}
\def\ii{{\mathbb{I}}}
\def\aa{{\cal A}}
\def\bb{{\cal B}}
\def\bbb{{\aa_I}}
\def\dd{{\cal D}}
\def\hh{{\cal H}}
\def\hhh{{\mathbb H}}
\def\jj{{\cal J}}
\def\oo{{\cal O}}
\def\ss{{\cal P}}
\def\mm{{\cal M}}

\def\lb{\left[}
\def\rb{\right]}
\def\lp{\left(}
\def\rp{\right)}
\def\pp{\pmatrix}

\def\L2{L_2(\mm)}
\def\LS{L_2(\mm , S)}

\def\cinf{C^{\infty}\lp\mm\rp}
\def\cm{C^{\infty}(\mm)}
\def\m3{M_3 \lp \cc \rp}
\def\m2{M_2 \lp \cc \rp}
\def\mx{M_k}
\def\mz{M_{k'}}
\def\mmp{M_p \lp \cc \rp}
\def\mn{M_n \lp \cc \rp}
\def\mnp{M_{np} \lp \cc \rp}
\def\mpn{M_{pn} \lp \cc \rp}
\def\cnp{\cc^{np}}
\def\cpn{\cc^{pn}}
\def\cn{\cc^{n}}
\def\cp{\cc^{p}}

\def\dm{\partial_{\mu}}
\def\dn{\partial_{\nu}}

\def\ot{\otimes}
\def\pof{\psi\otimesk}
\def\fop{k\otimes\psi}
\def\ds{\mbox{$\slash\!\!\!\partial$}}
\def\da{\left[ D,\pi(a) \right]}
\def\df{\left[ D,f \right]}

\def\oe{\omega_{E}}
\def\oi{\omega_{I}}
\def\oei{\omega_{E}\ot\omega_{I}}
\def\oeip{\omega_{E}\ot\omega_{I}'}
\def\oepi{\omega_{E}'\ot\omega_I }
\def\oepip{ {\omega_E}' \ot {\omega_I}'}
\def\xoi{\omega_x\ot\omega_i}
\def\xoip{\omega_x\ot\omega_{i}'}
\def\yoi{\omega_y\ot\omega_i}
\def\yoip{\omega_y\ot\omega_{i}'}
\def\ou{\omega_{1}}
\def\od{\omega_{2}}
\def\ox{\omega_{k}}
\def\oz{\omega_{{k'}}}
\def\oxe{\omega_{k_e}}
\def\oze{\omega_{{k'}_e}}
\def\oue{\omega_{1}\circ\alpha_e}
\def\ode{\omega_{2}\circ\alpha_e}
\def\o0{\omega_0}
\def\xox{x_{k}}
\def\yox{y_{k}}
\def\rx{\rho_k}
\def\rz{\rho_{k'}}
\def\ux{{u_k}}
\def\uz{{u_{k'}}}
\def\xoz{x_{{k'}}}
\def\yoz{y_{{k'}}}
\def\xo0{x_\omega^0}
\def\yo0{y_\omega^0}
\def\tox{\tilde{\omega_{k}}}
\def\toz{\tilde{\omega_{{k'}}}}
\def\nx{{n_k}}
\def\nz{{n_{k'}}}
\def\ae{\alpha_e}
\def\aea{\alpha_e(\aa)}
\def\be{\begin{equation}}
\def\ee{\end{equation}}
\def\beqn{\begin{eqnarray*}}
\def\eeqn{\end{\eqnarray*}}
\def\beqnn{\begin{\eqnarray}}
\def\eeqnn{\end{\eqnarray}}
\def\dm{\lp\begin{array}}
\def\fm{\end{array}\rp}

\def\dbb{\lb\begin{array}}
\def\fbb{\end{array}\rb}
\def\dbn{\left.\begin{array}}
\def\fbn{\end{array}\right.}

%\magnification=1200
\hsize 17truecm
\vsize 24truecm

\font\thirteen=cmbx10 at 13pt
\font\ten=cmbx10
\font\eight=cmr8

\baselineskip 15pt
%\baselineskip 20pt

%\renewcommand{\baselinestretch}{1.6}

%\nopagenumbers
\thispagestyle{empty}

%\centerline{\ten Centre de Physique
%Th\'eorique\footnote{\eight Unit\'e Propre de Recherche
%7061}, CNRS Luminy, Case 907}

%\centerline{\ten F-13288 Marseille -- Cedex 9}

$~$
\vskip 2truecm

%%*********** insert here your own title

\begin{center}
{\Large\bf DISCRETE KALUZA-KLEIN  FROM \\[1ex]
 SCALAR FLUCTUATIONS
\\[2.5ex]
IN NONCOMMUTATIVE GEOMETRY}

\vskip 2cm

{\bf Pierre MARTINETTI$^1$,  Raimar WULKENHAAR$^2$}

\vspace{4ex}

$^1$Centre de Physique Th\'eorique, CNRS Luminy, Case 907\\
F-13288 Marseille -- Cedex 9, France

\vskip 1ex

$^2$Institut f\"ur Theoretische Physik, Universit\"at Wien\\
Boltzmanngasse 5, A-1090 Wien, Austria

\footnotetext[1]{and Universit\'e de Provence, martinet@cpt.univ-mrs.fr}
\footnotetext[2]{Marie-Curie Fellow, raimar@doppler.thp.univie.ac.at}

\end{center}

\vskip 2truecm \centerline{\bf Abstract} \medskip
We compute the metric associated to noncommutative spaces described by
a tensor product of spectral triples. Well-known results of the
two-sheets model (distance on a sheet, distance between the sheets) are
extended to any product of two spectral triples. The distance between
different points on different fibres is investigated. When one of the
triples describes a manifold, one finds a Pythagorean theorem as soon
as the direct sum of the internal states (viewed as projections)
commutes with the internal Dirac operator. Scalar fluctuations yield
a discrete Kaluza-Klein model in which the extra component of the metric is
given by the internal part of the geometry. In the standard model, this
extra component comes from the Higgs field.  \vskip 2truecm

%%********** insert key-words if necessary
%%\noindent Key-Words: \\
%%PACS-92: 11.15 Gauge field theories \\
%%MSC-91: 81T13 Yang-Mills and other gauge theories
%%\bigskip
%
%********** insert number of figures
%\noindent Number of figures: 1
%\bigskip

%\noindent 20 mars 2001

%\noindent CPT-99/P.3860
%\bigskip

%\noindent anonymous ftp : ftp.cpt.univ-mrs.fr
%\noindent web : www.cpt.univ-mrs.fr

\newpage
\setcounter{page}{1}
\setcounter{footnote}{0}
\renewcommand{\thefootnote}{\alph{footnote}}

\section{Introduction.}

In the noncommutative approach to the standard model of elementary
particles\cite{gravity}, space-time appears as the product (in the
sense of fibre bundles) of a continuous manifold by a discrete space.
In precedent papers, we have studied the metric aspect of
several classes of discrete spaces\cite{finite}, and the metric of the
continuum has
been approached from a Lie-algebraic approach \cite{gyros}. Here,
within the framework of noncommutative geometry, we investigate how the
distance in
the continuum evolves when the space-time of euclidean general relativity is
tensorised by an internal space. We find that in many cases the relevant
picture is
the two-sheets model
\cite{connes,connesfrance}. Indeed, under precise conditions, the
metric aspect of "continuum $\times$ discrete" spaces reduces to the
simple picture of two copies of the manifold. It was
known\cite{metrique,cham2} that the distance on each copy is the geodesic
distance while the distance between the copies~-- the distance on the
fibre~-- is a constant. But this does not give a complete description of
the geometry, in particular the distance between different points on
different copies. In this
paper we show that this distance coincides with the geodesic distance within a
$(4{+}1)$-dimensional manifold whose fifth component comes from the internal
part of the geometry. This component is a constant in the simpliest cases and
becomes a function of the manifold when the metric fluctuates. Restricting
ourselves to scalar fluctuations of the metric, which correspond to the Higgs
sector in the standard model, it appears that the Higgs field describes the
internal part of the metric in terms of a discrete Kaluza-Klein model. 

The aim of this paper is to investigate the metric aspect of the
standard model geometry. This goal is only partially achieved because
we focus on scalar fluctuations and we mention only very briefly
mathematical aspects such as the Gromov distance. For a comprehensive
approach of
these questions, the reader is invited to consult ref.\cite{rieffel}. Other
works on
distance in noncommutative geometry mainly concern lattices
\cite{atzmon,bimonte,dimakis,manfred} and finite spaces. A larger
bibliography can be
found in ref.\cite{finite}. Naturally, using a Kaluza-Klein picture in
noncommutative geometry is not a new idea and one can refer to
refs.\cite{madore,coquereaux} for instance as well as the
textbook\cite{madore2}. Particularly, that the distance between the sheets
depends on the manifold has been shown in refs.\cite{cham1,cham2}.  
Last but not least, for a comprehensive approach of the subject, the most
recent and complete reference is the book \cite{jgb}.

The paper is written for a $4$-dimensional manifold but generalisation
to higher dimension should be straightforward. The next two sections
introduce classical notions of distance in noncommutative geometry and
a simple proof that, on a manifold, this distance coincides with the
geodesic distance.  Section IV extends known results of the two-sheets
model~-- distance on each copy, distance between the copies~-- to the
product of any two spaces (not necessarily a manifold $\times$ a
discrete space).  In section V we show that, under conditions on the
internal part of the Dirac operator, a large number of examples
actually reduce to a two points fibre space.  In the simplest case the
internal space is orthogonal to the continuum in the sense of
Pythagorean theorem (in finite spaces, the Pythagorean theorem has
already been mentioned by ref.\cite{pekin}).  Section VI studies the
scalar fluctuations (terminology is precised there) of this metric.
The last part presents examples, among them the standard model, and
precises the link between the Higgs field and the metric.

\section{The distance formula.}

Let $\aa$ be a unital $C^{*}$-algebra represented over a complex
Hilbert space
$\hh$ equipped with a scalar product $\scal{.}$ defining the norm
$\norm{\psi}_{\hh}^2\doteq\abs{\langle \psi,\psi \rangle}$ for $\psi
\in \hh$. The $C^{*}$\!-norm of $\aa$ is the operator norm in $\hh$
$$
\norm{a}_{\aa}\doteq \sup_{\psi\in\hh}
\frac{\norm{\pi(a)(\psi)}_{\hh}}{\norm{\psi}_{\hh}}
$$
where $\pi$ is the representation.  The so called Dirac operator
$D$ is a selfadjoint operator in $\hh$, possibly unbounded.  When the
spectral dimension is even \cite{connes}, the chirality $\chi$ is a
hermitean operator which anticommutes with $D$ and commutes with
$\pi(\aa)$. The set $(\aa, \hh, D, \pi, \chi)$ is called a spectral
triple. The terminology is justified because $\pi$ is usually infered
in the notation $\hh$, and once given $(\aa,\hh,D)$, $\chi$ -- if it
exists -- is uniquely determined by the axioms of noncommutative
geometry\cite{gravity}. Since the algebra appears through its
representation, we can, without loss of generality, replace $\aa$ by
$\aa\slash \text{ker}(\pi)$ and assume that $\pi$ is faithful. To
improve the readability we omit the symbol $\pi$ unless necessary.

We denote by $\ss(\aa)$ the set of pure states of $\aa$. The
distance $d$ between two of its elements $\omega_{1},\omega_{2}$ is
\begin{equation*}
d(\omega_{1},\omega_{2})\doteq \sup_{a\in\aa} \left\{ \,
\abs{\omega_{1}(a) -\omega_{2}(a)}\, /\, \norm{[D,a]}\leq 1
\right\}\,,
\end{equation*}
where $\norm{.}$ is the operator norm in $\hh$ (we do not write
$\norm{[D,a]}_\aa$ because $[D,a]$ may not be the representation of an
element of $\aa$).  This supremum is reached \cite{finite} by a
positive element such that $\norm{[D,\pi(a)]}=1$:
\begin{equation}
\label{distance}
d(\omega_{1},\omega_{2})= \sup_{a\in\aa_+} \left\{ \,
\abs{\omega_{1}(a) -\omega_{2}(a)}\, /\,
\norm{[D,a]}= 1 \right\} .
\end{equation}

This formula is invariant under several transformations, including
unitary transformation and projection.  First,
a unitary element $u$ of $\aa$ defines both an automorphism of the
algebra  $\alpha_u(a)\doteq uau^{*}$
and a {\it unitary equivalent} triple
$
(\aa,\hh, uDu^{*},\pi\circ\alpha_u)\,.
$ Obviously distances are not changed under such a transformation because
$\norm{[D,a]}=\norm{[uDu^{*}, \alpha_u(a)]}$.
More interesting is the
action of a projection $e\in\aa$ ($e^2=e^{*}=e$) through the endomorphism
of $\aa$
$$
\ae(a)\doteq eae,
$$
which defines the {\it restricted} spectral triple
$$
(\aa_e\doteq \aea,\quad \hh_e\doteq e\hh \doteq \text{ran}\,e ,\quad
D_e\doteq eDe\big|_{\hh_e},\quad
\pi_e\doteq \pi\big|_{\hh_e})
$$
whose corresponding distance is denoted by $d_e$. $\alpha_e$ being not
injective, for a pure state $\omega\in\ss(\aa)$ the linear form
$\omega\circ\alpha_e$ is not necessarily a state of $\aa$ (for instance if
$e$ is in the kernel of $\omega$). However it is a pure state of the
subalgebra $\aa_e$. Conversely, any pure state $\omega_{e}$ of
$\aa_e$ is made a pure state of $\aa$ by writing
$\omega_{e}\circ\alpha_e$. In other words,
$\ss(\aa_e)=\ss(\aa)\circ\alpha_e\subset\ss(\aa)$.

{\lem
\label{projection}
If a projection $e$ is such that $[D,e]=0$, the distance between
two pure states $\ou,\od$ of $\aa_e$ is invariant by projection:
$d_e(\ou,\od)=d(\oue,\ode)$.}
\newline

\noindent
{\it Proof.}
For
$a_e\in\aa_e$,
$\norm{[D_e,
\pi_e(a_e)]}=\norm{[\pi(e) D \pi(e), \pi(a_e)]}=
\norm{[ D, \pi(a_e)]}$ therefore
\begin{eqnarray*}
d_e(\ou,\od)
&=&\suup{a_e\in \aa_e} \left\{\abs{(\ou-\od)(a_e)}\, /\,
\norm{[D, \pi(a_e)]}\leq 1 \right\},\\
&\leq&\suup{a\in\aa} \left\{
\abs{(\oue-\ode)(a)}\, /
\,\norm{[D, \pi(a)]}\leq 1 \right\}= d(\oue,\ode).
\end{eqnarray*}
This upper bound is reached by $\alpha_e(a)$
where
$a\in \aa$ reaches the supremum for the distance $d$,
namely
$\norm{[D,\pi(a)]} = 1$ and
$d(\ou\circ\alpha_e, \od\circ\alpha_e)= \ou\circ\alpha_e (a) -
\od\circ\alpha_e (a)$.  \hfill$\blacksquare$

\section{ Distance in a manifold.}

The spectral triple of a Riemannian spin manifold $\mm$ of dimension
$4$ with a metric $g$ is
\begin{equation}
\label{tripletvariete}
\aa=\cm,\qquad \hh=\LS,\qquad
D=i\gamma^{\mu}\partial_{\mu}=i\ds\,,
\end{equation}
where $L_2(\mm,S)$ is the set of square integrable spinors on $\mm$.
The Riemannian gamma matrices $\gamma^\mu={\gamma^\mu}^*=e^\mu_a
\gamma^a$ are obtained via the vierbein field $e^\mu_a$ from the
Euclidean gamma matrices $\gamma^a$ of the associated Clifford
algebra. Using $\delta^{ab} e^\mu_a e^\nu_b = g^{\mu\nu}$ and $\gamma^a
\gamma^b + \gamma^b \gamma^a = 2 \delta^{ab}\ii$ one has $\gamma^\mu
\gamma^\nu + \gamma^\nu \gamma^\mu = 2 g^{\mu\nu}\ii$.
The spectral dimension is the dimension of the manifold, so there is a
chirality $\gamma^5 =\gamma^0\gamma^1\gamma^2\gamma^3$ made of the
Euclidean $\gamma^a$'s. The scalar
product of $\hh$ is $\scl{\psi}{\phi} \doteq\int_{\mm}
\bar{\psi}(x)\phi(x)\, dx$ and an element $f\in\aa$ is represented
over $\hh$ by the pointwise multiplication, $\pi(f)\doteq f\ii$, so
that
$$
\norm{f}_{\aa}=\sup_{\psi \in \hh}
\lp\frac{\int_{\mm}(\bar{f}\bar{\psi})(x)
(f\psi)(x)dx}{\int_{\mm}\bar{\psi}(x)\psi(x)dx}\rp^{\frac{1}{2}}
=\sup_{x\in\mm} \abs{f(x)}\,.
$$
By Gelfand transform, $\ss(\aa)\simeq \mm$. The isomorphism $\,
x\!\in\!\mm \leftrightarrow \omega_{x}\!\in\! \ss(\aa)\,$ is defined
by $\,\omega_x(f)\doteq f(x).$ The noncommutative
distance (\ref{distance})
\begin{equation*}
\label{distance1,5}
d(x,y)=\sup_{f\in\cinf} \left\{ \,
\abs{f(x) -f(y)}\, /\, \norm{[i\ds,f\ii]}\leq 1 \right\}\,,
\end{equation*}
coincides with the geodesic distance $L(x,y)$ between points $x,y$ of
$\mm$. This is a classical result \cite{connes} but the proof
introduces ideas and notations important for further presentation so
that we shall give it in detail (this version of the proof comes from ref.
\cite{krajew2}).

The supremum is reached on $\aa_+$,
so $f$ is real.  For $\psi\in\hh$, $[i\ds,f\ii]\psi=i(\ds f)\psi$, so
$$
\norm{[i\ds,f\ii]}^2=\norm{(i\ds f)^{*} i\ds f}
=\norm{\gamma^{\mu}\partial_{\mu}f\gamma^{\nu}\partial_{\nu}f}
=\norm{g^{\mu\nu}\partial_{\mu}f\partial_{\nu}f\ii}=
\sup_{q\in\mm} \left\{
g^{\mu\nu}(q)\partial_{\mu}f(q)\partial_{\nu}f(q)\right\}\,.
$$
The gradient $\grad$ in the usual sense is the exterior derivative
$d$ (not to be confused with the distance) which maps $0$-forms (i.e.\ 
smooth functions over $\mm$) onto $1$-forms:
\begin{equation*}
\label{gradient}
\grad f\doteq
(\partial_{\mu}f) dx^{\mu}\in T^{*}\mm\,.
\end{equation*}
By definition \cite{nakahara} $g$ defines an inner product (thus, a
norm) in each cotangent space $T^{*}_q\mm$ in such a manner that
$$
\norm{\grad{f(q)}}_{T^{*}_q\mm}^2=
%\scl{\partial_{\mu}f(q) dx^{\mu}}{\partial_{\nu}f(q)
%dx^{\nu}}_{T^{*}_q\mm}=\partial_{\mu}f(q)\partial_{\nu}f(q)\scl{
%dx^{\mu}}{ dx^{\nu}}_{T^{*}_q\mm}=
g^{\mu\nu}(q)\partial_{\mu}f(q)\partial_{\nu}f(q)\,.
$$
Omitting the index ${T_q}^*\mm$, one writes
$$
\norm{[i\ds,f\ii]}=\suup{q\in\mm}\norm{\grad{f}(q)}.
$$

Now,  let $c\!:\!t\in
[0,1]\!\!\rightarrow\!\!\mm$ be the minimal geodesic between $x$ and $y$ and
let $\dot{}$ denote the total derivative with respect to $t$. For any
$f\in\cinf$
$$f(x)-f(y) = \int_{0}^{1} \dot{f}(c(t))\; dt=\int_{0}^{1}
\partial_{\mu}f(p)\; \dot{c^{\mu}}(t) dt$$
where $p\doteq c(t)$.
The metric defines an isomorphism
$T_p\mm\simeq T^{*}_p\mm$
%\;(\dot{c}^{\mu}(t)\partial_{\mu}\!\mapsto\!\dot{c}_{\nu}(t)dx^{\nu}\!\doteq\!
%g_{\mu\nu}(p)\dot{c}^{\mu}(t )dx^{\nu}$)
such that
$$
{\partial_{\mu}f(p)\, \dot{c^{\mu}}(t)}={g^{\mu\nu}(p)\,
\partial_{\mu}f(p) \, \dot{c_{\nu}}(t)}={\scl{\grad
f(p)}{\dot{c}_{\nu}(t)dx^{\nu}}}\,,
$$
thus, by Cauchy-Schwarz, $\abs{{\partial_{\mu}f(p)\,
\dot{c}^{\mu}(t)}} \leq \norm { {\grad f}(p)} \norm{\dot{c}_\nu(t)
dx^\nu}$.  Assuming that $f$ reaches the supremum, one has
$\norm{\grad{f}}\leq1$, so
\begin{equation*}
\label{firstineq}
d(x,y) =\abs{f(x)-f(y)}\leq \int_0^1
\norm{\dot{c}_\nu(t)dx^\nu}\; dt = L(x,y)\,.
\end{equation*}

This upper bound is reached by the function
\begin{equation}
\label{defl}
L: q \mapsto L(q,y).
\end{equation}
Indeed, $L(x)-L(y) = L(x,y)$ and
\begin{equation}
\label{normedel}
\sup_{q\in\mm}\norm{{\grad L}(q)}\leq 1\,.
\end{equation}
To prove (\ref{normedel}), take $q,q'\in\mm$ with coordinates
$q^{\mu},{q'}^\mu$ in a given chart such that $q'$ comes from $q$ by
the infinitesimal transformation $\sigma(\epsilon)$, $\epsilon<\!\!<
1$, where $\sigma$ is the flow generated by the vector field
$g^{\mu\nu}(\partial_{\nu} L)\partial_{\mu}$ with initial condition
$\sigma(0)=q$. Then, writing $dq^{\mu}{\doteq q'}^{\mu}-q^{\mu}$,
$$
q^{\mu}+dq^{\mu}= {q'}^{\mu} = \sigma^{\mu}(\epsilon)
= \sigma^{\mu}(0) + \epsilon\, \frac{d \sigma^{\mu}}{dt}(0)
+ {\cal O}(\epsilon^2)
= q^{\mu} +\epsilon\, g^{\mu\nu}(q)\partial_{\nu} L(q)
+{\cal O}(\epsilon^2)\,,
$$
which means that
\begin{equation}
\label{flot}
dq^{\mu}=\epsilon\,g^{\mu\nu}(q)\partial_{\nu} L(q) +{\cal O}(\epsilon^2)\,.
\end{equation}
As $L(q',y)$ is the shortest length from $q'$ to $y$, $L(q',y)\leq
L(q',q) + L(q,y)$, and one has
\begin{equation}
\label{inegtrg} L(q+dq) \leq L(q',q) + L(q)\,.
\end{equation}
Using (\ref{flot}),
$$
L(q',q)\doteq \sqrt{g_{\lambda\rho}(q) dq^{\lambda}dq^{\rho}}=
\sqrt{\epsilon^2 g_{\lambda\rho}(q)  g^{\lambda\mu}(q) \partial_{\mu}L(q)\,
g^{\rho\nu}(q) \partial_{\nu} L (q)}= \epsilon \sqrt{
g^{\mu\nu}\partial_{\mu}L(q)\, \partial_{\nu} L (q)}\,.
$$
Inserting into the r.h.s. of (\ref{inegtrg}) whose l.h.s. is developed with
respect to $\epsilon$ yields
$$
L(q) + \partial_{\mu}  L(q) \, dq^{\mu}= L(q) + \,
\epsilon\,g^{\mu\nu}(q)\partial_{\mu}L(q)\partial_{\nu} L(q)
+{\cal O}(\epsilon^2)
\leq    \epsilon\sqrt{g^{\mu\nu}\partial_{\mu}L(q)\,  \partial_{\nu} L (q)}
+ L(q) +{\cal O}(\epsilon^2),
$$
which is true for all $q$, hence (\ref{normedel}) and finally
$d(x,y)=L(x,y)$.

\section{Tensor product of spectral triples.}

The tensor product of an even spectral triple $T_I=(\aa_I, \hh_I,
D_I,\pi_I)$ with chirality $\chi_I$ by the spectral triple
$T_E=(\aa_E, \hh_E, D_E,\pi_E)$ is the spectral triple $T_I\otimes
T_E\doteq(\aa',\hh',D')$ defined by
\begin{equation*}
\label{pdt0} \aa'\doteq \aa_I\otimes\aa_E,\quad
\hh'\doteq\hh_I\otimes\hh_E,\quad  D'\doteq D_I\otimes\ii_E
+ \chi_I\otimes D_E\,,
\end{equation*}
where the representation of $\aa'$ is $\pi'\doteq \pi_I\ot \pi_E$. The
notation $T_I\ot T_E$ is a matter of convention for spectral triples
do not form a vector space. The product of spectral triples is
commutative in the sense that when $T_E$ is even with chirality
$\chi_E$, then $T_E\otimes T_I\doteq(\aa,\hh,D)$ is well defined by
permutation of factors,
\begin{equation}
\label{pdt1}
\aa\doteq \aa_E\otimes\aa_I,\quad
\hh\doteq\hh_E\otimes\hh_I,\quad  D\doteq D_E\otimes\ii_I
+ \chi_E\otimes D_I\,,
\end{equation}
$\pi=\pi_E\ot\pi_I$, and is equivalent to $T_I\otimes T_E$ up to
the unitary operator
\begin{equation*}
%\label{U}
U\doteq (\frac{\ii_I+ \chi_I}{2}\otimes\ii_E +
\frac{\ii_I- \chi_I}{2}\otimes\chi_E)\,.
\end{equation*}

For physics it is interesting to take for this tensor product the product
of the continuum by the discrete, namely to study the
geometry of the four-dimensional space-time of Euclidean general
relativity together with an internal discrete space. In the standard
model, the internal space describes the electroweak and strong
interactions and is defined by a spectral triple $T_I$
in which the algebra $\aa_I$ is chosen such that its unitarities are related
to the gauge group of interactions while $\hh_I$ is the space of
fermions. Both $\aa_I$ and $\hh_I$ are finite dimensional, so $T_I$ is
a finite spectral triple\cite{krajew} and $T_E$ is the usual spectral
triple (\ref{tripletvariete}) of a manifold. The spectral dimension of
a finite spectral triple is $0$ and
$\text{dim}(T_E)=\text{dim}(\mm)=4$: both $T_E$ and $T_I$ are even
therefore both $T_E\ot T_I$ and $T_I\ot T_E$ are defined.

In this section,
we give general results
that do not require neither $T_E$ to be the spectral triple of a manifold
nor $T_I$ to be finite.
To fix notations we simply assume that $T_E$ is even so that we work with
$T_E\ot T_I$. To study the metric of a noncommutative space, the first goal is
to make  explicit the set of pure  states of the associated algebra. For $\oe$
and $\oi$ being pure states of $\aa_E$ and $\aa_I$, the pair $(\oe, \oi)$ is
a  state of $\aa$ which acts as $\omega_E\ot\omega_I$ (that
$\ii$ maps to $1$ is obvious, the positivity can be seen in
 ref.\cite{murphy} for instance) but this is not necessarily a pure
state.  Moreover there can be pure states  of $\aa$ that cannot be written
as tensor products. However, as soon as one of the algebras is abelian, one
obtains \cite{kadison} that
$\ss({\aa_E}\ot{\aa_I})
\simeq
\ss({\aa_E})
\times \ss({\aa_I})$ and any pure state $\omega$ of $\aa$ writes $\omega
=\oe\ot\oi$.

In the two sheets-model $\aa= \cinf\ot\cc^2$, therefore any
pure state is $\xoi$ where $\omega_i$,
$i=1,2$, is a pure state of $\cc^2$ and labels
the sheets. It is known\cite{connes} that
$d(\xoi,\, \yoi)$ is the geodesic distance $L(x,y)$
while   $d(\xoi,\, \omega_x\ot\omega_j)$ is a constant. This
 extends to any product of spectral triples. Once fixed a pure state
$\oe$, $d(\oei, \oeip)$ depends only on the spectral triple $T_I$ and,
similarly, $d(\oei,\oepi)$ depends only on $T_E$.
 This is true even when none of the algebra is commutative:  the distance is
then defined between states that may be not pure.

{\thm\label{sansfluct}
Let $d_E$, $d_I$, $d$ be the distance in $T_E$, $T_I$, $T_E\ot T_I$
respectively. For
$\oe$, $\oe'$ in $\ss(\aa_E)$ and $\oi$, $\oi'$ in $\ss(\aa_I)$,
\begin{eqnarray*}
&d\lp\oei, \oeip\rp= d_I\lp\oi,\oi'\rp,&\\
&d\lp\oei, \oepi\rp= d_E\lp\oe,\oe'\rp.&
\end{eqnarray*}}

\noindent
{\it Proof.} Let $f_j$ denote the elements of $\aa_E$ and $m_i$ those of
$\aa_I$.
A generic element of $\aa$ is
$
a=f^i\ot m_i\,,
$
where the summation index $i$ runs over a finite subset of $\nn$.
Definition (\ref{pdt1}) yields
$$
[D,a]= [D_E,f^i]\ot m_i + f^i\chi_E \ot [D_I,m_i]\,.
$$
Multiplying on left and right by the unitary operator $\chi_E\ot \ii_I$
allows to write
$$
\norm{[D_E, f^i]\ot m_i + f^i\chi_E\ot[D_I,m_i]}=\norm{ -[D_E,f^i] \ot m_i +
f^i\chi_E\ot[D_I,m_i]}\,,
$$
where we use that $\chi_E=\chi_E^*$ commutes with $f^i$ and anticommutes
with $D_E$.
For $u,v$ in a normed space,
$2\norm{u}\leq\norm{u+v}+\norm{u-v}$, thus
\begin{equation}
\label{gen1}
\norm{ [D_E, f^i]\ot m_i} \leq \norm{[D,a]} \,,
\end{equation}
and $\norm{f^i\chi_E\ot[D_I,m_i]} \leq  \norm{[D,a]}\,.$ One can factorise
the left-hand
side of this last equation by
$\chi_E\ot\ii_I$ in order to have
\begin{equation}
\label{gen2}
\norm{f^i\ot[D_I,m_i]} \leq  \norm{[D,a]}\,.
\end{equation}

For any $\oe\in \ss(\aa_E)$ and $a\in\aa_+$, let us define
$a_E\in\aa_I$ by
$$
a_E\doteq\oe (f^i)m_i.
$$
$a_E$ is selfadjoint. Indeed, positivity of $a$, i.e.\
$
a=({f^p}^{*}\ot {m_p}^{*})(f^q\ot m_q)= \frac{1}{2} ( f^{pq} \ot m_{pq} +
{f^{pq}}^{*}\ot {m_{pq}}^{*})
$
where $f^{pq}\doteq {f^p}^{*}f^q$ and $m_{pq}=m_p^{*}m_q$, yields
$$
a_E=\frac{1}{2}( \oe (f^{pq})m_{pq} +  \oe
({f^{pq}}^{*}){m_{pq}}^{*})= a_E^{*}.
$$
Thus
$$i[D_I,a_E]= i \lp \omega_E \ot \ii_I\rp \lp f^i\ot  [D_I,  m_I]\rp$$
in $\bb(\hh_I)$ is
normal. One knows\cite{kadison} that for any normal element $a$ of a
$C^*$-algebra,  $\norm{a}=\suup{\tau\in{\cal S}}\abs{\tau(a)}$, where $\cal
S$ is the set of states. Thus, with ${\cal S}_I$ the set of states of
$\bb(\hh_{I})$,
\begin{eqnarray*}
\norm{[D_I,a_E]} &  = & \suup{\tau_I\in\, {\cal S}_I} \abs{{\tau_I}
([D_I,a_E])},\\
                 &\leq&  \suup{({\tilde{\omega}}_E, \tau_I) \in \,
\ss(\aa_E)\times{\cal S}_I}\abs{( \tilde{{\omega}}_E\ot \tau_I) (f^i\ot
[D_I,m_i])},\\
&\leq&  \suup{(\tau_E, \tau_I) \in \,
{\cal S}_E\times{\cal S}_I} \abs { \lp \tau_E \ot \tau_I \rp \lp f^i\ot
[D_I,m_i]
\rp}= \norm{f^i\ot [D_I,m_i]},
\end{eqnarray*}
where we use that $if^i\ot [D_I,m_i]\in\bb(\hh)$ is also normal.
Together with (\ref{gen2}),
$$\norm{[D_I,a_E]}\leq \norm{[D,a]}.$$
Since $(\oei)(a)-(\oeip)(a)=\oi(a_E) - \oi'(a_E)$,
$$d(\oei,\oeip)\leq d_I(\oi,\oi').$$
This upper bound is reached by $\ii_E\ot a_I$ where $a_I\in\aa_I$
reaches the
supremum for $T_I$ alone, namely
$d_I(\oi,\oi')\doteq\abs{(\oi-\oi')(a_I)}$
and $1=\norm{[D_I,\pi_I(a_I)]}$.

The proof for $d(\oei,\oepi)$ is similar, using (\ref{gen1})
instead of (\ref{gen2}).
\hfill $\blacksquare$

\section{ Metric in the continuum $\times$ discrete.}

The key points of Theorem \ref{sansfluct} are equations
(\ref{gen1}) and (\ref{gen2}). The first one allows to forget about
the internal part of the commutator and makes sense for states
of $\aa$ defined by different pure states on $\aa_E$ but the
same pure state on $\aa_I$. When $T_E$ is the spectral triple
of a manifold and $T_I$ a finite spectral triple, the noncommutative
space described by $T_E\times T_I$ is a fibre
bundle over the manifold with a discrete fibre. This can also be seen
 as the union of several copies
of the manifold, indexed by the element of the fibre. Theorem
\ref{sansfluct} simply says that each of the copies
is endowed with the metric of the base. Note that
the discussion about the Gromov distance between manifolds with
distinct metrics in ref.\cite{connes} may not be transposed
here because such manifolds are not described by a tensor product
of spectral triples.

In contrast, (\ref{gen2}) does not take into account the external part
of the commutator and is sufficient to determine the distance between
states defined by the same pure state on $\aa_I$ (i.e.\ points on the
same fibre within the picture of a continuum $\times$ discrete space).
Of course the mixed case $d\lp \oei,\, \oepip\rp$ ~-- the
distance between different points on different copies of the
manifold~-- requires to take into account both the internal and the
external part of the commutator. This makes the computation more
difficult. However, for continuum $\times$ discrete spaces, some of
these distances have a nice interpretation in terms of a discrete
Kaluza-Klein model: although the internal space is discrete, the
distance appears as the geodesic distance in a "virtual"
$(4{+}1)$-dimensional manifold ("virtual" means that the points
between the sheets are not part of the geometry, the embedding into a
higher dimensional continuum space is a practical intermediate).

Let us first give a semi-general result which does not require $T_E$ to be the
spectral triple of a manifold
 ($T_E$ is just supposed to be even to fix notations) but which assumes
\begin{equation*}
\label{sommeinterne}
\aa_I\doteq\bigoplus_{k} \aa_k\,,
\end{equation*}
where $k$ runs over a finite subset of $\nn$ and the $\aa_k$'s are von
Neumann algebras -- i.e. their
universal representation $\{ \pi_u, \hh_u \}$ is a von Neumann
algebra --  on $\cc$. Note that
a pure state of a direct sum of algebras is a pure state of one of the
algebras, that is
$$\ss(A_I)=\underset{k}{\bigcup}\, \ss(\aa_k).$$
The reason why we restrict to von Neumann algebras is that to
any pure states $\omega$ of $\aa_k$ corresponds a projection $\rho\in\aa_k$
such that
\begin{equation}
\label{projecteur}
\alpha_\rho(a)\doteq \rho a \rho = \omega(a) \rho.
\end{equation}
This result comes from the proof of proposition 2.16 of ref.\cite{takesaki}
in which is assumed, by hypothesis, that the universal enveloping von Neumann
algebra $\tilde{\aa_k}$ equals $\pi_u(\aa_k)$. Strictly speaking
this proof is written for complex algebras. However in the
standard model, we shall explicitly exhibit such a  projection  for the
real internal algebra so that, in the following, we deal with algebra over
$\kk$ where $\kk=\cc$ or $\rr$. Typically, in physical examples, the
$\aa_k$ are matrix algebras and
$\rho$ is a density matrix. When pure states of different components $\aa_k$
are involved and
$D_I$ commutes with the direct sum of the corresponding projectors,  one
obtains as an immediate
consequence of Lemma \ref{projection} that $\aa_I$ reduces to $\kk^2$.

 {\prop \label{reduction} For $\omega_k\in \ss(\aa_k)$,
let $\rho$ be the corresponding projection in $\aa_k$. Define
similarly $\rho'$ for $k'\neq k$ and let $p\doteq \rho \oplus
\rho'$.
If $[D_I, p]=0$ then, for any $\oe,\oe'\in\ss(\aa_E)$,
$$d\lp \oe\ot\omega_k,\, \oe'\ot\omega_{k'}\rp =d_e\lp \oe\ot\ou,\,
\oe'\ot\od\rp$$
where $\ou, \od$ are the pure states of $\,\kk^2$ and $d_e$ is the distance
associated
to
$T_e\doteq T_E\ot T_r$ with $$ \aa_r\doteq\kk^2,\quad \hh_r\doteq
 p \hh_I,\quad
D_r\doteq p D_I p  \big|_{\hh_r}.$$}
\noindent
{\it Proof.}
The projection $e \doteq \ii_E \ot p \in\aa$ defines the restricted triple
$T_e\doteq(\aa_e,
\hh_e, D_e)$ in which
$$
\aa_e\doteq \alpha_e(\aa)  = \aa_E \ot \alpha_{p} \lp\aa_I\rp.
$$
Since $\rho$ and $\rho'$ correspond to different components of $\aa_I$
they are
orthogonal, therefore
$$\alpha_p(\aa_I)=\alpha_{\rho}\lp\aa_k\rp\oplus \alpha_{\rho'}\lp
{\aa_{k'}}\rp
=\omega_k(\aa_k)\rho \oplus {{\omega_{k'}}}({\aa_{k'}})\rho'
$$
 by (\ref{projecteur}). $\omega_k, \omega_{k'}$ being surjective on $\kk$,
$\omega_k(\aa_k)\rho$ and $\omega_{k'}({\aa_{k'}})\rho'$ are isomorphic to
$\kk$. Hence
\begin{equation*}
\label{algebrer}
\aa_e = \aa_E\ot \kk^2.
\end{equation*}
The state $\omega_i\in\ss(\kk^2)$ extracts the $i^{th}$  component of a pair of
elements of $\kk$. In
detail, for
$a_I\in\aa_I$,
\begin{equation}
\label{a0}
\alpha_{p} (a_I) = \omega_k(a_I)\rho \, \oplus \,  \omega_{k'} (a_I)\rho'\,,
\end{equation} so that
$\omega_1\circ \alpha_{p} (a_I)= \omega_k(a_I).$
Since $e$ acts like the identity on $\aa_E$,
$$ (\oe \ot \ou)\circ\alpha_e = \oe \ot  (\ou\circ\alpha_{p})= \oe\ot
\omega_k,$$ and
$(\oe'\ot\od)\circ\alpha_e = \oe'\ot\omega_{k'}$. By hypothesis $[ D  , e]=
\chi_E\ot [ D_I  , p]= 0$ so Lemma
\ref{projection} yields
$$d\lp \oe\ot\omega_k,\, \oe'\ot\omega_{k'}\rp =d_e\lp \oe\ot\ou,\,
\oe'\ot\od\rp.$$
$\hh_r$ and  $D_r$ are given by Lemma \ref{projection}. \hfill$\blacksquare$
\newline

To explicitly compute $d_e$, we now focus on the case of a
continuum $\times$ discrete space and we take for $T_E$ the spectral
triple of a manifold (\ref{tripletvariete}). To simplify the notations,
the pure state
$\omega_x\ot\omega_k$ is denoted by
$\xox$. The main result of this section is that the internal space is
orthogonal  to the manifold, in
the sense
of Pythagorean theorem, as soon as the Dirac operator commutes with the sum of
the density matrices.

{\thm
\label{pythagore}
Let $\ox,\oz\in\ss(\aa_k),\ss(\aa_{k'})$, $k\neq k'$. Let $\rho, \rho'$
be the associated projections and $p\doteq \rho\oplus \rho'$. If
$[D_I,p]=0$, then
for any points $x,y$ in
$\mm$
$$d(x_k, y_{k'})^2 = d(x_k, y_k)^2 + d(y_{k}, y_{k'})^2.$$}
\noindent
{\it Proof.} The proof consists of three steps. First the problem is
reduced to a two-sheets model.  Then the distance is shown to be the
geodesic distance within a $(4{+}1)$-dimensional Riemannian manifold
which, third, satisfies Pythagorean theorem.

1) With notations of Proposition \ref{reduction},
\begin{equation}
\label{etapeun}
d(x_k, y_{k'})= d_e(x_1, y_2).
\end{equation} Let us be more explicit on $\hh_r, \pi_r$ and
$D_r$.
\begin{equation}
\label{h0}
\hh_r \doteq p\hh_I = \hh_k \oplus \hh_{k'}, 
\end{equation}
where $\hh_k \doteq \rho \hh_I$ and $\hh_{k'}\doteq \rho'\hh_I$. Following
(\ref{a0}), one lets
$a_r =\omega_k(a_I)\rho \oplus
\omega_{k'}(a_I) \rho'$
denote a generic element of $\aa_r$. Clearly $\pi_r(\rho)=\ii_k$ so
\begin{equation}
\label{pi0}
\pi_r(a_r) = \omega_k(a_I) \ii_k \oplus
\omega_{k'}(a_I)\ii_{{k'}}.
\end{equation}
$D_r$ is the restriction to $\hh_r$ of the projection of $D_I$ on $\hh_r$,
namely 
\begin{equation}
\label{dio}
D_r\doteq\dm{cc} V&M\\ M^{*}& W \fm
\end{equation}
where $M$ is a linear application from $\hh_k$ to $\hh_{k'}$, 
and $V$, $W$ are endomorphisms of $\hh_k$, $\hh_{k'}$ respectively. $M$ is
supposed to be non zero for the contrary makes $D_r$ commuting
with $\pi_r$, that is all states  of $\aa$ defined by $\omega_k$ are at
infinite distance from any
states defined by $\omega_{k'}$.

Equations (\ref{h0}, \ref{pi0}, \ref{dio}) associated to (\ref{a0})
fully determine the triple
$T_r$, and thus $T_e$. Omitting  $\rho$ and $\rho'$ appearing in (\ref{a0}),
a generic element of $\aa_e$
writes
$$a = f^i \ot \omega_k (m_i) \, \oplus \, f^i \ot \omega_{k'}(m_i) = f
\oplus g,
$$
where $ m_i\in \aa_I$ and $f^i, f\doteq f^i\omega_k(m_i), g\doteq
f^i\omega_{k'}(m_i)
\in \cinf$.
In accordance with (\ref{distance}), we assume that $f\oplus g$ is positive,
i.e.\ $f$ and
$g$ are real functions. $x_1$ and $y_2$ act as
$$x_1(a) =f(x), \qquad y_2(a) = g(y).$$
$a$ is represented by
$$f\ii_E\ot \ii_{k} \, \oplus \, g\ii_E \ot \ii_{{k'}}$$
and the Dirac operator $D_e= i\ds \ot \ii_I + \gamma^5 \ot D_r$ is such that
\begin{equation}
\label{co}
[D_e, a] =\dm{cc}
i\ds f \ot\ii_{k} & (g-f)\gamma^5\ot M \\[1ex]
\overline{(f-g)}\gamma^5\ot M^{*}
& i\ds
\ot\ii_{{k'}}\fm.
\end{equation}

2) Let us show that $d_e$ coincides with the geodesic distance on the
compact manifold
$$
\mm'\doteq [0,1] \times \mm\,,
$$
with coordinates ${x'}^a=(t, x^\mu)$, equipped with the metric
\begin{equation*}
\{g^{ab}(x')\}\doteq\dm{cc} \norm{M}^2 &0\\ 0 & g^{\mu\nu}(x)\fm\,,
\end{equation*}
and made a spin manifold by adding to the previous $\gamma$-matrices
$$
\gamma^t=\norm{M} \gamma^5.
$$
Thanks to section III, it is
enough to show that $d_e$ coincides with the distance $L'$ of the triple
$$
\aa'=C^{\infty}(\mm'),\qquad \hh'=L_2(\mm',S), \qquad
D'=i\gamma^a\partial_a=i\gamma^t\partial_t+ i\ds\,.
$$

To proceed, let $\aa''$ be the subset of ${\aa'}_+$  consisting of all
functions
$$
\phi (t,x)\doteq (1-t) f(x) + tg(x),
$$
where $f$ and $g$ are any real functions on $\mm$. Then
\begin{eqnarray*}
\nonumber
\norm{[D',\phi]}^2 &=& \norm{\gamma^a \partial_a \phi}^2
= \sup_{(t,x)\in\mm'} \, \left\{
g^{ab}(t,x)\,\partial_a\phi(t,x)\,
\partial_b\phi(t,x) \right\}
\\
           &\leq&\label{ameq3} \sup_{x\in\mm} \left\{ \,
\abs{(f-g)(x)}^2\norm{M}^2  + \suup{t\in[0,1]}\,  P(t,x)\, \right\},
\end{eqnarray*}
where
$$
P(t,x)\doteq t^2 \norm{\grad{(f-g)(x)}}^2 + 2tg^{\mu\nu}(x)\,\partial_\mu
(f-g)(x)\,\partial_\nu g(x)+\norm{\grad{g(x)}}^2
$$
is a parabola in $t$ of positive leading coefficient, i.e.\  which reaches
its maximum for
$t=0$ or $1$. Note that
$$P(0,x)= \norm{\grad{g(x)}}^2,\qquad  P(1,x)=\norm{\grad{f(x)}}^2$$
and, thanks to (\ref{co}),
\begin{eqnarray*}
 &\norm{ \dm{cc} \ii_E \ot \ii_{k} &~0 \\ 0&~0 \fm  [D_e,a]\dm{cc}
\ii_E\ot
\ii_{k}&0
\\ 0 & \gamma^5\ot\ii_{{k'}}\fm }^2=\norm{
\dm{cc} i\ds f\ot \ii_{k} ~& (g-f)\ii_E\ot M\\ 0 ~& 0\fm}^2 &
\\
&=\suup{x\in\mm}\;\left\{\,\norm{\grad{f(x)}}^2 +\, \abs{f(x)-g(x)}^2
\norm{M}^2\,
\right\} \leq
\norm{[D_e,a]}^2.&
\end{eqnarray*}
Similarly, one has
$\;\suup{x\in\mm}\;\left\{\,\norm{\grad{g(x)}}^2 +\, \abs{f(x)-g(x)}^2
\norm{M}^2\,
\right\}\leq \norm{[D_e,a]}^2$, hence
$$
\norm{[D',\phi]}\leq \norm{[D_e,a]}.
$$
Consequently, since $x_1(a) - y_2(a) = \phi(0,x)-\phi(1,y)$,
\begin{equation}
\label{etapedeux}
d_e(x_1, y_2)\leq \sup_{\phi\in{\aa''}} \left\{
\abs{\phi(0,x)-\phi(1,y)}\,/\,\norm{[D',\phi]\leq1} \right\} \leq
L'\lp(0,x),(1,y)\rp\,.
\end{equation}

Proving the converse inequality calls for more precisions on the geometry
of $\mm'$. Because $\{g^{ab}(x')\}$ is block diagonal and does not depend
on $t$,
 the coefficients of the Levi-Civita connexion are
\begin{eqnarray*}
\label{gamma}
& \Gamma^t_{t\mu}=\Gamma^t_{\mu t}=
\frac{1}{2}g^{tt}\partial_{\mu} g_{tt}\,,\qquad
\Gamma^\mu_{tt}=-\frac{1}{2} g^{\mu\nu} \partial_\nu g_{tt}\,, &
\\
&\Gamma^{\mu}_{t\nu}=\Gamma^{\mu}_{\nu t} =
\Gamma^{t}_{tt}=\Gamma^t_{\mu\nu}=0\,, &
\end{eqnarray*}
where $g_{tt}=(g^{tt})^{-1}=\norm{M}^{-2}.$ The geodesic equations read
\begin{eqnarray}
&\displaystyle
\frac{d^2 t}{d\tau^2} + g^{tt}(\partial_\mu g_{tt})
\frac{dt}{d\tau}\frac{dx^{\mu}}{d\tau}= 0\,,&
\label{geo1}
\\
\label{geo2}
&\displaystyle
\frac{d^2 x^\mu}{d\tau^2} - \frac{1}{2} g^{\mu\nu}(\partial_\nu g_{tt})
\frac{dt}{d\tau}\frac{dt}{d\tau} +\Gamma^{\mu}_{\lambda\rho}\frac{d
x^\lambda}{d\tau}\frac{d x^\rho}{d\tau} = 0\,,&
\end{eqnarray}
and, because $g_{tt}$ does not depend on $x^\mu$, reduce to
\begin{equation}
\label{constante}
\frac{dt}{d\tau}= \text{constant}\doteq g^{tt}K\; \text{ and }\;\,
\frac{d^2 x^\mu}{d\tau^2}
+\Gamma^{\mu}_{\lambda\rho}\frac{dx^\lambda}{d\tau}\frac{d x^\rho}{d\tau} =
0\,,
\end{equation}
where $K$ is a real constant. In other terms, the projection to $\mm$ of
a geodesic $\cal{G}'$ of $\mm'$ is a geodesic $\cal{G}$ of $\mm$, and
the projection of $\cal{G'}$ to the submanifold $[0,1] \times {\cal G}$ is
a straight line (i.e.\ a geodesic of the submanifold).
Let $\{x^a (\tau)\}$ be a geodesic in $\mm'$ parametrised by its length
element $d\tau$. Note that, using (\ref{constante}),
\begin{equation}
\label{dessin}
1= \frac{d\tau^2}{d\tau^2} =
g_{\mu\nu}\frac{dx^\mu}{d\tau}\frac{dx^\nu}{d\tau}+ g^{tt}K^2.
\end{equation}
 Let  $ds$ be the line element of $\mm$. Assuming that $g^{tt}K^2\neq1$ (this
will be discussed later),
\begin{equation}
\label{dtau}
d\tau^2 =  \frac{ds^2}{1 - g^{tt}K^2},\quad
dt = \frac{dt}{d\tau}d\tau = \frac{g^{tt}Kds}{\sqrt{1-g^{tt}K^2}}.
\end{equation}
For $q$ in $\mm$, let ${\cal G}_q'$ be the minimum geodesic of $\mm'$ between
$(0,q)$ and
$(1,y)$, and ${\cal G}_q$ its projection on $\mm$. Let us define
$f_0\in\cinf$ by
$$
\label{gun}
f_0(q) = \sqrt{1-g^{tt}K^2} L(q) = \sqrt{1-g^{tt}K^2}\int_{{\cal G}_q}ds,$$
where $L$ has been defined in (\ref{defl}).
Take $a_0=(f_0,g_0)\in\aa_e$, where $g_0=f_0 - K$. Then
\begin{equation}
\label{fg}
x_1(a_0)-y_2(a_0) = f_0(x) - g_0(y)= f_0(x) + K.
\end{equation}
But the second equation (\ref{dtau}) gives
$$1=\int_{{\cal G}_x'} dt =  \frac{g^{tt}K}{\sqrt{1-g^{tt}K^2}}\int_{{\cal
G}_x}  ds,
$$
inserted in (\ref{fg}) as $K1$,
$$
x_1(a_0)-y_2(a_0)=\sqrt{1-g^{tt}K^2}\int_{{\cal G}_x}
ds+\frac{g^{tt}K^2}{\sqrt{1-g^{tt}K^2}}\int_{{\cal G}_x} ds =
\frac{1}{\sqrt{1-g^{tt}K^2}}\int_{{\cal
G}_x}ds.$$
Using the first equation (\ref{dtau}) one obtains
\begin{equation}
x_1(a_0)-y_2(a_0) = \int_{{\cal G}_x'} d\tau=  L'\lp(0,x),(1,y)\rp.
\label{presque}
\end{equation}
Moreover, $\ds f_0=\ds g_0$ and
$\partial_\mu f_0 = \sqrt{1 - g^{tt}K^2}\partial_\mu L$,  so (\ref{co}) yields
\begin{eqnarray*}
\norm{[D_e, a_0]}^2 &=&\suup{q\in\mm} \left\{g^{\mu\nu}(q)\partial_\mu f_0(q)
\partial_\nu f_0(q) +
g^{tt}K^2\right\}\\
& =& \suup{q\in\mm} \left\{ (1 - g^{tt}K^2)\norm{\grad{L}(q)}^2 +
g^{tt}K^2\right\}.
\end{eqnarray*}
Recalling (\ref{normedel}), this gives $\norm{[D_e, a_0]}\leq 1$ so, with
(\ref{presque}),
$$d_e(x_1, y_2)\geq L'((0,x),(1,y)).$$
Together with (\ref{etapedeux}) and (\ref{etapeun}),
\begin{equation}
\label{resultat1}
d(\xox,\yoz)= L'((0,x),(1,y))\,.
\end{equation}
This result holds as long as $g^{tt}K^2\neq 1$. If this is not true,
 then $U\doteq \frac{dx^\mu}{d\tau}\partial_\mu\in TM$ is zero for
(\ref{dessin})
indicates that
$g(U,U)=0$ and $\mm$ is Riemannian. In other words, $x^{\mu}(\tau)$ is a
constant.
This cannot be the equation of ${\cal G'}_x$ unless $x=y$. As a
conclusion, (\ref{resultat1}) holds as soon as $x\neq y$.

When $x=y$, (\ref{etapeun}) gives $d(\yox,\yoz)= d_e(y_1,y_2)$. With
$d_r$ denoting the distance associated to the triple $T_r$ alone,
Proposition (\ref{sansfluct}) yields $d_e(y_1,y_2)=d_{r}(\ou, \od),$
which is nothing but the distance of the simplest two-points space and
equals\cite{connes} $\frac{1}{\norm{M}}$. Thus
\begin{equation}
\label{resultat3}
d(\yox,\yoz)=\frac{1}{\norm{M}}.
\end{equation}
The projection ${\cal G}_y$ of the geodesic ${\cal G'}_x={\cal G}'_y$ is,
by (\ref{geo2}), a geodesic between $y$ and $y$, that is to say
a point. ${\cal G}'_y$ reduces to a straight line in the hyperplane.
Thus $d\tau^2=g_{tt}dt^2$ and
$$L'\lp (0,y),(1,y)\rp= \sqrt{g_{tt}}\int_{{\cal G}_y'} dt
=\sqrt{g_{tt}}=\frac{1}{\norm{M}}.$$
Consequently
$
d(\yox,\yoz)=L'\lp (0,y),(1,y)\rp
$
and (\ref{resultat1}) holds even if $x=y$.
\bigskip

3)  The last step is to show that (\ref{resultat1}) satisfies Pythagorean
equality.
$g^{tt}$ being a
constant, equation (\ref{dtau}) indicates that $d\tau$ and $ds$ are equal
up to
a constant factor. In this
way, one may parametrise a geodesic of $\mm'$ by $ds$ rather than $d\tau$
and obtains,
thanks to the
geodesic equations,
$$dt= g^{tt}K'ds$$
where $K'$ is a real constant. Then
$$d\tau^2 = g_{tt}dt^2 + ds^2 =ds^2(1 + g^{tt}{K'}^2).$$
Thus
\begin{eqnarray}
\nonumber
L'\lp(0,x),(1,y)\rp &=& \sqrt{1 + g^{tt}{K'}^2}\int_{{\cal G}'_x} ds= \sqrt{1
+ g^{tt}{K'}^2}L(x,y)\\
\label{pythun}
&=&\sqrt{L(x,y)^2 + g^{tt}{K'}^2L(x,y)^2}.
\end{eqnarray}
On one side, Theorem \ref{sansfluct} gives $L(x,y)=d(\xox,\yox)$. On
the other side,
$$g^{tt}{K'}^2L(x,y)^2= g_{tt}\lp\int_{{\cal G}_x'} g^{tt}{K'} ds\rp ^2=
g_{tt}\lp\int_{{\cal G}_x'} dt \rp
^2=g_{tt}=\frac{1}{\norm{M}^2}=d^2(\yox,\yoz)$$
by (\ref{resultat3}). Together with (\ref{resultat1}) and (\ref{pythun}),
\begin{equation*}
d(\xox,\yoz)^2=d(\xox,\yox)^2 + d^2(\yox,\yoz)\,.
\tag*{\mbox{$\blacksquare$}}
\end{equation*}

\section{Fluctuations of the metric.}

For a complete presentation of the material of this section and a
justification of the terminology, see
refs.\cite{connes,gravity}.  To a triple
$(\aa,\hh,D)$, the axiom of reality adds an operator $J$, called
the real structure, such that $[JaJ^{-1},b]=0$ for any
$a,b\in\aa$.  This allows to define a right action of $\aa$ over $\hh$
which makes sense because of the noncommutativity of the algebra. To
define a notion of unitarily equivalent spectral triples preserving the
operator $J$,
a unitary element $u$ of $\aa$ is implemented by the operator
$U\doteq uJuJ^{-1}$ rather than the operator $u$. Then
the action of $u$ defines the {\it gauge transformed} triple
$(\aa,\hh, D_A)$ where
\begin{equation}
\label{da}
D_A\doteq UDU^{*}= D+A+ JAJ^{-1}
\end{equation}
with
$$
A\doteq u[D,u^{-1}]\,.
$$
The selfadjoint operator $A$ governs the failure of invariance of $D$ under
a gauge transformation \cite{connes}. Under a
gauge transformation, $A$ transforms like a usual vector potential.
Since in electrodynamics the vector potential is a 1-form, one defines
the space $\Omega^1$ of 1-form of the noncommutative space
$(\aa,\hh,D)$ as the set of elements
$$a^i[D,b_i]$$ where $a^i,b_i\in\aa$. Note that  we use the
simplifying notation $\Omega^1$ rather than $\Omega_D^1$, more common in the
literature, because we  only deal with 0-forms and 1-forms ($\Omega^n_D$
differs from
$\Omega^n$ for $n\geq 2$). Since $A$ is selfadjoint, the set of
vector potentials is simply the subset of selfadjoint elements of
$\Omega^1$. For any vector potential $A$, $D_A$ defined by (\ref{da}) is
called the covariant Dirac operator.

The distance is not invariant under a gauge transformation
and the metric is said to fluctuate. To study such fluctuations,
one has to replace $D$ by $D_A$ everywhere in the preceding sections. A
well known result makes this replacement less studious than it seems.

{\lem \label{trace1forme}
$[a, J\omega J^{-1}]=0,\; \forall \omega\in\Omega^1,a\in\aa$.}
\newline

\noindent
{\it Proof.}
$[J^{-1}aJ,[D,b_i]]=0$ (first order axiom)  and
$[a,J a^iJ^{-1}]=0$ (axiom of reality) yield
\begin{align*}
[a, J\omega J^{-1}]&= [a,J a^i[D,b_i]J^{-1}] \\
                        &= aJ a^iJ^{-1}J[D,b_i]J^{-1}-Ja^i[D,b_i]J^{-1}a \\
                        &= J a^i[D,b_i]J^{-1}a - Ja^i[D,b_i]J^{-1}a=0 \,.
\tag*{\mbox{$\blacksquare$}}
\end{align*}
As an immediate consequence,
\begin{equation}
\label{ca}
[D_A,a]=[D+ A,a]\,.
\end{equation}
Let us now work out the 1-forms of a tensor product triple $T_E\ot
T_I$. In refs. \cite{kt,schucker} it is shown that
$$
\Omega^1=\Omega^1_E\ot \Omega^0_I + \chi_E\Omega^0_E\ot\Omega^1_I\,,
$$
where $\Omega^0_E=\aa_E$ is the set of 0-forms of $\aa_E$, and similar
definitions for the other terms. When $T_E$ is the spectral triple of a
manifold,
$$
\Omega_E^1\ni f^j[i\ds,g_j\ii_E]= if^j(\gamma^\mu \partial_\mu
g_j)=i\gamma^\mu f_\mu\,,
$$
where $f^j,g_j,f_\mu\doteq f^j\partial_\mu g_j \in\cinf$.
A 1-form of the total
spectral triple is
$$
\Omega^1\ni i\gamma^\mu f_\mu^i \ot a_i + \gamma^5 h^j\ot m_j$$
where $a_i\in\aa_I$, $h^j\in\cinf$, $m_j\in\Omega_I^1$. A vector potential is
\begin{equation}
\label{h}
A = i\gamma^\mu \ot A_\mu + \gamma^5\ot H
\end{equation}
with $A_\mu\doteq {f^i}_\mu a_i$ an $\aa_I$-valued skew-adjoint vector
field (over $\mm$) and $H\doteq h^j m_j$ an
$\Omega^1_I$-valued selfadjoint scalar field.
For a matrix algebra (or a direct sum of matrix algebras), the
skew-adjoint elements form the Lie algebra of the Lie group of
unitarities. This Lie group represents the gauge group of the theory, thus
$A_\mu$ is a gauge potential. In ref.\cite{gravity} a formula is given for the
fluctuations of the
metrics due to $A_\mu$. Here, we focus on the fluctuations coming from
the scalar field $H$ only, and we assume that $A_\mu=0$. Then (\ref{ca})
becomes
\begin{equation}
\label{daa}
[D_A,a]= [D + \gamma^5\ot H,a].
\end{equation}

{}From now on, we write $D_A\doteq D + \gamma^5\ot H$. For simplicity,
$d$ still denotes the distance associated to the triple $(\aa, \hh,
D_A)$. Remembering definition (\ref{pdt1}), a scalar fluctuation
substitutes
$$D_H\doteq D_I+ H$$
for $D_I$. The main difference is that the internal Dirac operator $D_H$
now depends on
$x$ so that each point $x$ of the manifold defines an internal triple
$$
T_I^x\doteq (\aa_I, \hh_I, D_H(x))\,.
$$
This interpretation of scalar fluctuations perfectly fits to the adaptation of
Theorem \ref{sansfluct}.
\newline
\newline
\noindent
{\it {\bf Theorem 2'.}
\label{fluct}
Let $L$ be the geodesic distance in $\mm$ and $d_x$ the distance of the
spectral
triple $T_I^x$ alone.
For $x,y \in \mm$ and $\omega_k, \omega_{k'}\in \ss(\aa_I)$,
\begin{eqnarray*}
d(\xox,\xoz ) &=& d_x(\omega_k,\omega_{k'}),\\
d(\xox,\yox) &=& L(x,y).
\end{eqnarray*}}

\noindent
{\it Proof.} The adaptation of the proof of Theorem \ref{sansfluct} is
straightforward. Notations
are similar except that $\omega_E$ is now $\omega_x$ so that $a_E$ is replaced
by $a_x$. With $H= h^j m_j$,
\begin{eqnarray}
\nonumber
[D_H(x), a_x] &=& [D_I + \omega_x(h^j) m_j, \omega_x(f^i)m_i ]\\
\nonumber
              &=& \omega_x(f^i)[D_I, m_i] + \omega_x(h^j)\omega_x(f^i) [
m_j, m_i ]\\
\label{remarque}
              &=& ( \omega_x \ot \ii_I ) \lp f^i\ot [D_I, m_i] + h^j f^i
\ot [  m_j, m_i ]\rp\\
\nonumber
              &=& ( \omega_x \ot \ii_I ) \lp f^i\ot [D_H, m_i ]\rp.
\end{eqnarray}
Then, $i[D_H(x), a_x]$ being normal,
\begin{eqnarray*}
\norm{[D_H(x), a_x]} &=& \suup{\tau_I\in {\cal S}_I}\abs{
\tau_I\lp[D_H(x),a_x]\rp}\\
                     &=& \suup{\tau_I\in {\cal S}_I}\abs{
(\omega_x\ot\tau_I)\lp
f^i\ot [D_H, m_i]\rp}\\
                  &\leq& \suup{\tilde{\omega}_E\ot \tau_I\in \ss(\aa_E)\ot
{\cal
S}_I}
\abs{ (\tilde{\omega}_E\ot\omega_I)\lp
f^i\ot [D_H, m_i ]\rp}\\
                  &\leq& \norm{ f^i\ot [D_H, m_i] }.
\end{eqnarray*}
Equation (\ref{gen2}) being replaced by $\norm{f^i \ot [D_H, m_i]}\leq
\norm{[D_A,a]}$,
one obtains
$$\norm{[D_H(x), a_x]}\leq\norm{[D_A,a]}.$$
The rest of the proof is then similar as in Theorem \ref{sansfluct}.
\hfill $\blacksquare$
\newline
\newline
Note that in (\ref{remarque}) we use that $\omega_x$ is a character, i.e.\
that $\aa_E$ is Abelian. 

Applied to the two sheets-model, Theorem \ref{sansfluct}
simply says that the distance between the sheets is encoded by a scalar field,
as it has already been shown in ref.\cite{cham1} (see also ref.\cite{cham2}
for a $M_2(\cc)\oplus\cc$ model). Theorem \ref{pythagore} is modified in a
more serious way for the fluctuation introduces an $x$-dependence for the
coefficients of the Kaluza-Klein metric. 
\newline

\noindent
{\it {\bf Theorem \ref{pythagore}'.}  Let
$\ox,\oz\in\ss(\aa_k),\ss(\aa_{k'})$,
$k\neq k'$. Let $\rho,
\rho'$ be the associated projections and $p\doteq \rho\oplus \rho'$.
If $[D_H, p]=0$ for any points of
$\mm$, then for any points
$x,y\in\mm$,
$$d(\xox,\yoz)= L'((0,x),(1,y))$$
where $L'$ is the geodesic distance of the spin
manifold $\mm'\doteq [0,1]\times \mm$ equipped with the metric
$$\dm{cc} \norm{M(x)}^2 & 0 \\ 0 & g^{\mu\nu}(x) \fm$$
in which $g^{\mu\nu}$ is the metric of $\mm$ and $M$ is the restriction to the
representation of $\aa_{k'}$ of the projection of $D_H$ on the
representation of
$\aa_{k}$.}
\newline

\noindent
{\it Proof.} Unless otherwise made precise, notations are similar to Theorem
\ref{pythagore}. The first
part of the proof is hardly modified. Let
$\psi^r\ot\xi_r\in\hh$. Recalling (\ref{daa}) and the definition
(\ref{h}) of $H$,
$$
[D_A, a] \psi^r\ot\xi_r = \gamma^5 \psi_r \ot [D_I, p] \xi_r + \gamma^5
h^j\psi_r
\ot [m_j,p]\xi_r\in
\hh.$$
Evaluated at $x\in\mm$, the above expression yields
$$[D_A, a] \psi^r (x) \ot\xi_r = \gamma^5 \psi_r(x) \ot [D_I +  H(x),
p]\xi_r = 0
$$
by hypothesis, which means that $[D_A , a]$ is the zero endomorphism of $\hh$
so that Lemma
\ref{reduction} applies and
$$d(\xox,\yoz)=d_e(x_1,y_2).$$
The only difference with Theorem \ref{pythagore} is that
$D_r$ now depends on $x$. For instance when $\aa_I$ is finite dimensional then
$M$ is a matrix whose entries are scalar fields on $\mm$.

Now $g^{tt}(x)\doteq\norm{M(x)}^2$ depends on
$x$ but is still independent with respect to $t$. The
geodesic equations (\ref{geo1}, \ref{geo2}) no longer reduce to
(\ref{constante}) but
\begin{eqnarray*}
\frac{d}{d\tau}
(g_{tt}\frac{dt}{d\tau})&=&(\frac{d}{d\tau}g_{tt})\frac{dt}{d\tau}+
g_{tt}\frac{d}{d\tau}(\frac{dt}{d\tau})\\
                                        &=&(\partial_\mu
g_{tt})\frac{dt}{d\tau}\frac{dx^\mu}{d\tau}+ g_{tt}\frac{d^2t}{d\tau^2}\\
                                        &=&g_{tt}\lp g^{tt}(\partial_\mu
g_{tt})\frac{dt}{d\tau}\frac{dx^\mu}{d\tau}+ \frac{d^2t}{d\tau^2}\rp=0
 \end{eqnarray*}
by (\ref{geo1}). Thus $g_{tt}\frac{dt}{d\tau}=K$ is a constant. This is almost
the first equation
(\ref{constante}), except that
\begin{equation}
\label{constanteh}
\frac{dt}{d\tau}=Kg^{tt}(x)
\end{equation}
now depends on $x$. $a_0= (f_0, g_0)$ is defined by
\begin{equation}
\label{gunh}
f_0(q) \doteq
%\int_{{\cal G}_q'}
%\sqrt{1-K^2g^{tt}}\frac{ds}{d\tau}d\tau=
\int_{{\cal G}_q}\sqrt{1-K^2g^{tt}}ds\,,\qquad
g_0 \doteq f_0 -K\,,
\end{equation}
where ${\cal G}_q'$ is the minimal geodesics from $(0,q)$ to the fixed point
$(1,y)$ and ${\cal G}_q$
its projection to $\mm$ (note that ${\cal G}_q$ is no longer a geodesic of
$\mm$).
Assuming that
\begin{equation}
\label{restriction3}
K^2g^{tt}(p)\neq 1
\end{equation}
for any $p \in {\cal G}_q$ allows to write
$d\tau = \frac{ds}{\sqrt{ 1 - K^2 g^{tt}} }$
and then
\begin{equation}
\label{dtauh}
1=\int_{{\cal G}_q'} dt = \int_{{\cal G}_q'} \frac{dt}{d\tau} d\tau =
\int_{{\cal G}_q}
\frac{Kg^{tt}}{\sqrt{1-K^2g^{tt}}}ds\,.
\end{equation}
 If (\ref{restriction3}) does not hold, we call $G$ the set of points $p$
of ${\cal G}_q$ for which $1 - K^2g^{tt}(p)=0$. $G'$ is the
corresponding subset of ${\cal G}'_q$. For any $p'\in {\cal G'}_q$,
(\ref{constanteh}) yields
\begin{eqnarray*}
%&\displaystyle
%\frac{dt}{d\tau}(q')d\tau= \frac{Kg^{tt}(p)ds}{\sqrt{1-K^2g^{tt}(p)}}\;
%\text { for } p\in {\cal G}_x/G\,, &
%\\
%&\displaystyle
\frac{dt}{d\tau}d\tau=K^{-1}d\tau\,,&
\end{eqnarray*}
and (\ref{dtauh}) is replaced by
$$
1 = \int_{{\cal G}_q/G}
\frac{Kg^{tt}}{\sqrt{1-K^2g^{tt}}}ds+\int_{G'}K^{-1} d\tau\,.
$$
Inserted as $K 1$ in $x_1(a_0) -y_2(a_0) = f_0(x) + K$, this gives
\begin{eqnarray*}
\nonumber
x_1(a_0) -y_2(a_0)
&=&\int_{{\cal G}_x} \sqrt{1-K^2g^{tt}} ds + \int_{{\cal G}_x/G}
\frac{K^2g^{tt}}{\sqrt{1-K^2g^{tt}(x)}}ds
+ \int_{G'} d\tau \\
&=& \int_{G}
\sqrt{1-K^2g^{tt}}ds  + \int_{{\cal G}_x/G}
\frac{ds}{\sqrt{1-K^2g^{tt}(x)}}  +
\int_{G'} d\tau \\ &=&\int_{{\cal G}_x'/G'} d\tau + \int_{G'} d\tau
= L'\lp(0,x),(1,y)\rp.
\end{eqnarray*}
The function $f_0(q)$ is in the vicinity of $q$ by definition
(\ref{gunh}) constant on a codimension 1 hypersurface through
$q$. Choosing an adapted reference frame with $\{x^1,x^2,x^3\}$ being the
coordinates in the hypersurface and $x^0$ the normal coordinate, one has
$ds(q) = \sqrt{g_{00}(q)} dx^0$ and $\partial_\mu f_0(q)=\delta_\mu^0
\partial_0 f_0 (q)$, giving
\begin{eqnarray*}
\partial_\mu f_0(q) &=& \delta_\mu^0 \sqrt{1-K^2g^{tt}(q)}\sqrt{g_{00}(q)}\,,
\\
g^{\mu\nu}(q)\partial_\mu f_0(q)\partial_\nu f_0(q) &=&
g^{00} (1 - g^{tt}K^2) g_{00}
= 1 - g^{tt}K^2 \,,
\end{eqnarray*}
which leads to
$\norm{[D_e,a_0]}=1$. Hence the result. \hfill $\blacksquare$
\newline

Few comments about this theorem. First, since all the coefficients of
the metric depend on $x$, there is no way that the geodesic distance
satisfies Pythagorean theorem. Second, a metric is non-degenerate by
definition, and we implicitly assume that $M(x)$ never cancels. This
was assumed in Theorem \ref{pythagore} to make the distance finite.
Here the point is more subtle for the field $M$ may be zero for some
points $x$. Let $\ker(M)\subset\mm$ be the set of such points. For any
$q\in \ker(M)$, $d\lp (0,q),(1,q)\rp=+\infty$ by Proposition 2'.
Moreover,
\begin{eqnarray*}
d \lp (0,q),(1,q)\rp &\leq&  d\lp (0,q),(0,x) \rp + d \lp
(0,x),(1,y)\rp + d\lp (1,y),(1,q) \rp\\
                       &\leq&   L(p,x) + d\lp (0,x),(1,y) \rp + L(y,q)\,,
\end{eqnarray*}
so $d\lp (0,x),(1,y)\rp=+\infty$ for any $x,y\in\mm$, which
contradicts Theorem 4' if $x=y\notin \ker(M)$. One
solution is to assume that any point $(t,q)$ with $q\in \ker(M)$ is at
infinite distance
from any other point, and define $\mm'$ as $[0,1]\times\mm/\ker(M)$.
If any path between $x$ and $y$ crosses $\ker(M)$, this operation
splits $\mm'$ into disconnected parts. A better solution is to take
into account the non-scalar part $A_\mu$ of the
fluctuation\footnote{In physical models, $M(x)$ is the representation
  of the Higgs field in the unbroken phase. Then, at $M=0$ the Higgs
  potential reaches its local maximum. Neglecting the gauge potential
  $A_\mu$, the Faddeev-Popov determinant of the t'Hooft gauge-fixing
  condition is zero at the maximum of the Higgs potential. This leads
  to a Gribov problem and questions a quantum treatment of $M(x)$
  without gauge field. (observation by Helmuth H\"uffel)}.  This goes
beyond the aim of this paper and the reader should refer to
ref.\cite{gravity}. 

\section{The standard model and other examples.}

We shall investigate the metric of spaces whose internal part is one of
those described in ref.\cite{finite}. We also give some indications on the
distance in the standard model.

\subsection*{Commutative spaces.}

We call commutative space a spectral triple whose internal algebra
is $\cc^k$, $k\in\nn$. Any $k$-tuple of complex numbers $a=(a^1,...,a^k)$ is
represented by
a diagonal matrix. For two pure states $\omega_u, \omega_v$ ($u,v\in [1,k]$),
$\rho_u\oplus \rho_v$ is the matrix with null coefficients except $1$ on the
$u^{th}$ and $v^{th}$ elements of the diagonal. Within the graphical
framework of ref.\cite{finite}, one shows that the internal distance only
depends on points that are on some path between $u$ and $v$. In other
terms
$$d_I(u,v) = \tilde{d}_I (u,v)$$
where $\tilde{d}_I$ denotes the distance computed with the Dirac operator
$\tilde{D}_I=\rho D_I \rho$ in which
$$\rho\doteq \underset{i\in\,  P \cup Q}\bigoplus\rho_i,$$ 
 with $P$ the set of points that are not connected to $u$ nor $v$, and $Q$ the
set of points that are connected to $u$ or \-- this is an exclusive "or"\--
$v$ by one and only one path. Note that, for any internal $1$-form,
$$\rho\,  a_i \,[D_I, b^i] \rho = a_i [ \tilde{D}, b^i]$$
so that the $\tilde{}$ notation is
coherent with the scalar fluctuation. At any point $x$ of the
manifold
$$d_x(\omega_u, \omega_v)= \tilde{d}_x(\omega_u, \omega_v)
$$
therefore, to  apply Theorem \ref{sansfluct}, it is enough to check that
$[\tilde{D}_H,
\rho]=0$. One verifies that whenever a component of the internal Dirac
operator is zero, the corresponding component of any internal 1-form is also
zero, so that $[\tilde{D}_H,\rho_I]=0$ as soon as $[\tilde{D}_I, \rho_I]=0$.

This means that the only path between $u$ and $v$ is the link
$u-v$ itself. The simplest case, $k=2$, endows the two-sheets model with a
cylindrical metric. The other examples of commutative spaces given in
ref.\cite{finite} do not fit the required condition and our next
examples will be noncommutative.

\subsection*{ Two-points space.}

Let $\aa_I=M_n(\cc)\oplus \cc$ be represented over $\cc^{n + 1}$ by
\begin{equation}
\label{rep2points}
\dm{cc} m~& ~0\\ 0~& ~c\fm
\end{equation}
where $m\in M_n(\cc)$ and $c\in\cc$. Possible chirality $K$ and Dirac
operator $\Delta$ are
$$K=\dm{cc}\ii_n&0\\0&-1\fm,\quad \Delta=\dm{cc}
0& M\\ M^{*}&0 \fm,
$$
where $M\in\cc^n$. But there is no operator $J$ to
fluctuate the metric. A solution is to make (\ref{rep2points}) acting over
$\hh_I = M_{n+1}(\cc)$ and define
$$
D\psi\doteq\Delta\psi + \psi\Delta,\qquad \chi_I\psi\doteq K\psi +
\psi K,\qquad J\psi \doteq \psi^{*}
$$
for any $\psi\in\hh_I.$ Since $J\Delta J^{-1}\psi= J\Delta\psi^{*}=
(\Delta\psi^{*})^{*}=\psi\Delta$, one has $D\psi= \Delta\psi + J\Delta
J^{-1}\psi$.
Moreover, for any $a\in\aa_I$, $[J\Delta
J^{-1},a]\psi=a\psi\Delta-a\psi\Delta=0$, so
$[D_I,a]=[\Delta,a]\,.$ Note that this result comes directly from Lemma
\ref{trace1forme} as soon
as one knows that $\Delta$ is a 1-form\cite{krajew}. Since the operator
norm over $\cc^n$ is equal to the operator norm over $M_n(\cc)$,
$$\norm{[D,a]}= \norm{[\Delta,a]}
$$
and the distance is in fact the same as the one computed with the spectral
triple
$(\aa_I,\cc^{n+1},\Delta)$. Note that this point is assumed in
ref.\cite{rovelli}.

Let $\rho_1$ be the density matrix associated to a pure state $\omega_1$ of
$\mn$ and
$\rho_0$ the one corresponding to the pure state $\omega_0$ of $\cc$. Then
$$\rho_1 \oplus \rho_0 = \dm{cc} \rho_1 & ~0 \\ 0& ~1 \fm$$
so that $[D_I, \rho_1 \oplus \rho_0]=0 $ is equivalent to
$\rho_1 M = M.$ In other terms, $M$ is colinear to the range of $\rho_1$.
 An happy coincidence makes that this is precisely the
condition under which the internal distance $d_I(\omega_1, \omega_0)=
\frac{ 1}{\norm{M}}
$ is finite\cite{finite}. Theorem 4 is true for any Dirac
operator~-- $d_I(\omega_0, \omega_1 )= + \infty$ makes $d(x_0,
y_1)=+\infty$ for
any $x,y$ in $\mm$~--  so
$$d (x_0, y_1) = \sqrt{L(x,y)^2 + \frac{1}{\norm{M}^2}  } $$
when $M$ is in the range of $\rho_1$, is infinite otherwise.

\subsection*{ The standard model.}

The spectral triple of the standard model (see
refs.\cite{connes,gravity,spectral} and ref.\cite{bridge,cham3} for a
physical expectation of the Higgs mass) is the tensor product of the usual
spectral triple of a manifold $T_E$ by an internal triple in which
$$\aa_I=\hhh \oplus \cc \oplus M_3(\cc)$$
(${\hhh}$ is the real algebra of quaternions) is represented over
$$\hh_I=\cc^{90}=\hh^P \oplus \hh^A= \hh_L^P \oplus \hh_R^P \oplus \hh_L^A
\oplus \hh_R^A\,.$$
The basis of $\hh_L^P=\cc^{24}$ consists of the left-handed fermions
$$
\dm{c} u\\d \fm_L,\;\dm{c} c\\s \fm_L,\;\dm{c} t\\b \fm_L,\;\dm{c}
\nu_e\\e \fm_L,\;\dm{c} \nu_\mu\\\mu \fm_L,\;\dm{c} \mu_\tau\\ \tau
\fm_L,
$$
 and the basis of $\hh_R^P=\cc^{21}$ is labelled by the right-handed fermions
$u_R,\, d_r,\, c_R,\, s_R,\, t_R,\, b_R \text{ and }\, e_R,\, \mu_R,\,
\tau_R$ (the model assumes massless neutrinos). The colour index for the
quarks has been omitted. $\hh_R^A$ and $\hh_L^A$ correspond to the
antiparticles. $(a\in\hhh,\; b\in\cc,
c\in M_3(\cc))$ is represented by
\begin{equation*}
\label{repms}
\pi_I(a,b,c)\doteq \pi^P (a,b) \oplus \pi^A (b,c) \doteq \pi_L^P(a) \oplus
\pi_R^P(b) \oplus \pi_L^A(b,c) \oplus \pi_R^A (b,c)
\end{equation*}
where, writing $B\doteq\dm{cc} b~&~0\\0~&~\bar{b}\fm\in\hhh$ and $N = 3$ for
the number of fermion generations,
\begin{eqnarray*}
\pi_L^P(a)\doteq a\ot\ii_N\ot\ii_3\,  \oplus\, a\ot\ii_N\,,\qquad& &
\pi_R^P(b) \doteq B \ot \ii_N \ot \ii_3\, \oplus \, \bar{b}\ot \ii_N\,,\\
\pi_L^A(b,c)\doteq \ii_2\ot\ii_N\ot c \,\oplus\,
\bar{b}\ii_2\ot\ii_N\,,\qquad & &
\pi_R^A(b,c)\doteq\ii_2\ot\ii_N\ot c \, \oplus \, \bar{b}\ii_n\,.
\end{eqnarray*}
One defines a real structure
$$
J_I= \dm{cc} 0~ & ~\ii_{15N}\\ \ii_{15N}~ & ~0\fm \circ\, \bar{}
$$
where $\,\bar{}\,$ denotes the complex conjugation, and an internal
Dirac operator
$$
D_I\doteq\dm{cc} D_P & 0\\ 0& \;\bar{D_P} \fm= \dm{cc} D_P~ & 0 \\ 0&0\fm +
J_I \dm{cc} D_P~ & 0 \\ 0&0\fm J^{-1}_I$$
whose diagonal blocks are $15N\times 15N$ matrices
$$D_P \doteq \dm{cc} 0&M\\ M^{*}& 0\fm,
$$
with  $M$ a $8N \times 7N$ matrix
\def\masseproj{\lp e_{11} \ot M_u + e_{22} \ot M_d \rp}
\def\masse{ \dm{cc}\masseproj \ot \ii_3& 0 \\0&e_2\ot M_e\fm}
\begin{equation}
\label{m}
M\doteq\masse.
\end{equation}
Here, $\{e_{ij}\}$ and $\{e_i\}$ denote the canonical basis of $\m2$ and
$\cc^2$ respectively. $M_u$, $M_d$, $M_e$ are the mass matrices
$$M_u=\dm{ccc} m_u & 0&0 \\ 0&m_c&0\\ 0&0&m_t\fm,\quad
M_d=C_{KM}\dm{ccc} m_d
& 0&0 \\ 0&m_s&0\\ 0&0&m_b\fm
,\quad M_e=\dm{ccc} m_e & 0&0 \\ 0&m_\mu&0\\ 0&0&m_\tau\fm
$$
 whose coefficients are the masses of the elementary fermions,
pondered by the unitary Cabibbo-Kobayashi-Maskawa matrix. The chirality,
last element of the spectral triple, is
$$
\chi_I= (-\ii_{8N}) \oplus \ii_{7N} \oplus (-\ii_{8N}) \oplus
\ii_{7N}\,.
$$

The presence of the conjugate representation $\bar{b}$ in $\pi_I$
requires to view $\cc$ as a real algebra. Therefore, the pure state
$\omega_0$ of $\cc$ is no longer the identity but an $\rr$-linear function with
value in $\rr$ which maps $1$ to $1$. In other words, $\omega_0$ is
the real part: $\omega_0(b)=\text{Re}(b)$.  As a quaternionic algebra,
$\hhh$ has a single pure state and this remains true for $\hhh$ seen
as a real algebra.

{\lem
\label{etatquaternion}
%Let $a\in\hhh$ be represented over $\cc^2$ by
%$\,\dm{cc} \theta~ & ~\rho \\-\bar{\rho}~& ~\bar{\theta} \fm$.
The single pure state $\omega_1$ of $\hhh$ is
$\omega_1(a)= \frac{1}{2}\text{Tr} (\ii_{\hhh}\,  a)$.}
\newline

\noindent
{\it Proof.}  The representation of $\hhh$ over the  four-dimensional real
vector space with basis $\{1,i,j,k\}$ such that $i^2=j^2=k^2=-1$,
$ij=-ji=k$, $jk=-kj=i$ and $ki=-ik=j$, is
$$a=\alpha + \beta i + \gamma j + \delta k$$
where $\alpha,\beta,\gamma, \delta\in\rr.$
Since $\bar{a}\doteq\alpha-\beta i - \gamma j - \delta k$,
$a\bar{a}\in\rr^+$ so any $\rr$-linear form is positive. Therefore a state
is any $\rr$-linear form that maps $\ii_\hhh=1$ to $1$.  Let
$\omega$ be such a state. By linearity,
$$\omega(a)=\alpha + \beta\omega(i) + \gamma\omega(j) + \delta\omega(k),$$
so $\omega$ is uniquely determined by its values on $i,j,k$. Let
$\omega_{\omega(i)}$ be the linear form defined by
$\omega_{\omega(i)}(i)=\omega(i)$,
$\omega_{\omega(i)}(1)=\omega_{\omega(i)}(j)=\omega_{\omega(i)}(k)= 0$.
Define similarly
$\omega_{\omega(1)}, \omega_{\omega(j)}, \omega_{\omega(k)}$. Then
\begin{eqnarray}
\nonumber
\omega &=& \omega_{\omega(1)} + \omega_{\omega(i)} +  \omega_{\omega(j)} +
\omega_{\omega(k)}\\
\label{quat}
       &=& \lambda (\omega_{\omega(1)} + \omega_{\kappa\omega(i)} +
\omega_{\kappa\omega(j)} +  \omega_{\kappa\omega(k)} )+ (1-\lambda)
(\omega_{\omega(1)} + \omega_{\kappa'\omega(i)} +  \omega_{\kappa'\omega(j)} +
\omega_{\kappa'\omega(k)}),
\end{eqnarray}
where $\lambda,\kappa\in\rr/\{1\}$ and $\kappa'\doteq
\frac{1-\lambda\kappa}{1-\lambda}$. Both factors of the right hand
side of (\ref{quat}) map $1$ to $1$, so they are states and $\omega$ is not
pure
 unless $\omega(i)=\omega(j)=\omega(k)=0$. Hence the only pure state of $\hhh$
is $\omega_1\doteq\omega_{\omega(1)}.$

The quaternion $a$ can also be represented over $\cc^2$
by $\dm{cc} \theta & ~\rho \\ -\bar{\rho} &~\bar{\theta}\fm$ where
$\theta\doteq\alpha + i \beta$. Then $\text{Tr} (a) \doteq 2 \text {Re}
(\theta) = 2 \alpha = 2\omega_1(a),$ that is $\omega_1(a)= \text{ Tr}
(\frac{1}{2}\ii_H \,  a)$.\hfill
$\blacksquare$
\newline

\noindent
 With regard to
$\ss(M_3(\cc))$, we shall only need the following well-known lemma:

{\lem
\label{etatm3}
Let $\omega,\omega'\in\ss(\aa_I).$ Then $\omega=
\omega'$ if and only if $\ker(\omega)=\ker(\omega')$.}
\newline

\noindent
{\it Proof.} Pure states are linear form, so if they have the same kernel
they are proportional. Since they coincide on the identity, they are
equal. \hfill $\blacksquare$
\newline

Noncommutative geometry gives an interpretation of the Higgs field as
a 1-form of the internal space. By
scalar fluctuation, 1-forms closely interfere with the metric. Thus
the Higgs field has an interpretation
in term of an internal metric. The conclusive result of this paper is
a precision of this link between Higgs and metric when the gauge field
$A_\mu$ is neglected.

{\prop
\label{distancems}
The finite part of the geometry of the standard model with scalar
fluctuations of the metric consists of a two-sheets model labelled by
the single states of $\cc$ and $\hhh$. Each of the sheets is a copy of
the Riemannian  four-dimensional space-time endowed with its metric. The
fifth component of the metric, corresponding to the discrete
dimension, is
$$
g^{tt}(x)= \lp\abs{1+h_1(x)}^2+\abs{h_2(x)}^2\rp m_t^2
$$
where $\dm{c} h_1\\h_2\fm$ is the Higgs doublet and $m_t$ the mass of the
quark top.}
\newline

\noindent
{\it Proof.} $\pi_I$ stands for $\pi_I(a,b,c)$ and $\Delta\doteq\dm{cc}
D_P~&0\\0&0\fm$ so that $D_I=\Delta+ J\Delta J^{-1}$. Since $\Delta$ is a
1-form\cite{krajew}, Lemma \ref{trace1forme} yields  $[J_I\Delta J^{-1}_I,
\pi_I] =0$, so that we can take $D_H= \Delta + H$. By explicit
calculation\cite{detail},
\begin{equation*}
\label{hms}
H=\dm{cccc}0&\pi^P_L(h)M~&~0~&~0~\\
M^{*}\pi^P_L(h^{*})&0&0&0\\0&0&0&0\\0&0&0&0\fm
\end{equation*}
where $h$ is a quaternion-valued scalar field.  Thus
\begin{equation}
\label{massems}
D_H = \dm{cccc}0&\Phi M~&~0~&~0~\\
M^{*}\Phi^{*}&0&0&0\\0&0&0&0\\0&0&0&0\fm,
\end{equation}
where
\begin{equation*}
\label{Phi}
\Phi\doteq (h + \ii_\hhh )\ot\ii_{4 N}=\dm{cc} 1 + h_1~ & ~h_2\\
-\bar{h}_2~& ~1+ \bar{h}_1\fm\, \ot \ii_{ 4N},
\end{equation*}
with $h_1$ and $h_2$ being two complex scalar fields.

By (\ref{distance}), the metric of the standard model is identical to the
metric associated to the triple $ (\aa_s,\, \hh, \,D)$, where
$\aa_s = {\cinf}_s \ot {\aa_I}_s$ is the subalgebra of selfadjoint elements of
$\aa$, with
$${\aa_I}_s = \cc_s \oplus \hhh_s \oplus M_3(\cc)_s =  \rr\,
\oplus\,\rr\oplus M_3(\cc)_s.$$
The representation $\pi_s$ associated to this triple coincides with the
restriction of $\pi$ to $\aa_s$. Concerning the quaternion, $\pi_s$
substitutes
$$\dm{cc} \theta & ~0\\ 0 & ~\theta \fm \; \text{ to } \;
\dm{cc} \theta &~\bar{\rho} \\ -\bar{\rho} &
~\bar{\theta}\fm.$$ 
In other words, to each representation of $\hhh$ there 
corresponds the direct sum of twice the fundamental representation of $\rr =
\hhh_s$. Now $\omega_1$ seen as a pure state of $\hhh_s$ is nothing but
the identity. The associated projection  $\rho_1\in\hhh_s$ is nothing but the
real number
$1$ which obviously satisfies (\ref{projecteur}). The same is true for
$\omega_0$ seen as a pure state of
$\rr=\cc_s$. Hence
$$\pi_s(\rho_0 \oplus \rho_1) = \dm{cc} \ii_{15 N} & 0 \\ 0&
\dm{cccc} 0_{6N} &       &        & \\
                 & \ii_{2 N} &        & \\
                 &       &0_{6N}& \\
                 &       &        &\ii_{N}
\fm
\fm
$$
commutes with $D_H$ defined in (\ref{massems}). Proposition 4'
applies to the distance between pure states of $\aa$ defined by $\omega_0$ and
$\omega_1$. Here
$$
\pi_s( \text{ran } \rho_1) = \hh^P_L\, \text{ and }\,
\pi_s(\text{ran }\rho_0) = \hh^P_R \oplus \hh^A_{lep}\, ,
$$
where $\hh^A_{lep}= \cc^{3 N}$ is the subset of $\hh^A$ generated by the
anti-leptons. Thus the extra metric component is
$$
g^{tt}(x) = \norm{\Phi(x) M}^2.
$$
Note that, as desired,  $\Phi M$ is a $2 \alpha^\hhh \times \lp \alpha^\cc +
\alpha^{\bar{\cc}}\, \rp$ matrix,  where $\alpha^\hhh= 4N$ is the degeneracy
of the representation of $\hhh_s$ in $\pi^P_L$, and
$\alpha^\cc=3N$, $\alpha^{\bar{\cc}}= 4 N$ are defined as well. Using the
explicit form (\ref{m}),
\begin{eqnarray*}
\norm{\Phi(x) M}^2 &=& \max \left\{ \,\norm{ (\Phi(x)\ot\ii_3)
( e_{11}\ot M_u + e_{22}\ot M_d)}^2,
\norm{(\Phi(x)\ot\ii_3)(e_2\ot M_e)}^2\, \right\}\\
&=& \lp\abs{1+h_1(x)}^2+\abs{h_2(x)}^2 \rp \max \left\{ \, {m_t}^2 ,
{m_\tau}^2
\right\}\\
&=& \lp \abs{1+h_1(x)}^2+\abs{h_2(x)}^2 \rp {m_t}^2.
\end{eqnarray*}

The other distances, involving the pure states of $M_3(\cc)$, are not
finite.  Indeed,
$$\norm{[D_H, \pi_I(a, b, c)]}= \norm{
\Big[\dm{cc} 0 & \Phi M \\ M^* \Phi^* & 0 \fm, \pi^P(a,b)\Big]}$$
does not put any constraint on $c$, thus for $\omega_2\in\ss\lp
M_3(\cc)\rp$ and
$\omega\in\ss(\aa_I)$,
$$
d_I(\omega_2,\omega) \geq \suup{c \in M_3(\cc)} \abs{\omega_2(c) -
\omega_(c)}.
$$
For $\omega=\omega_0$, $c=\lambda\ii_3$ with $\lambda\rightarrow\infty$ makes
the distance $d_I(\omega_2, \omega_0)$ infinite. Then
$$
d_I(\omega_2, \omega_0) = d(x_2, x_0) \leq d (x_2, y_0) + d(y_0, x_0)
\leq d (x_2, y_0) + L(x,y)
$$
by Theorem 4', so that $d (x_2, y_0) = +\infty $. The same is true
for $\omega=\omega_1$. The same is also true when
$\omega\in\ss(M_3(\cc))$ because, by
Lemma \ref{etatm3}, there exists
$c'\in ker({\omega_2})$,  $c'\notin ker(\omega)$ which makes $d_I(\omega_2,
\omega)$ infinite.  \hfill $\blacksquare$

\section{Conclusion.}

Noncommutative geometry intrinsically links the Higgs field with the
metric structure of space-time. We have not considered the gauge field
$A_\mu$ so it is not clear whether or not the interpretation of the
Higgs as an extra metric component has a direct physical meaning. It
is important to study the influence of the gauge fluctuation and,
particularly, how it probably makes the metric of the strong
interaction part finite.

Since $\hhh$ has only one pure state, the problem of the distance between
states defined by distinct pure states of the same component of the
internal algebra is not questioned here. One may be tempted to
consider states
$\tau$ of $\hhh$ that are not pure. But asking
$\tau(\bar{q})=\bar{\tau}(q)$ \-- which is part of the definition of a real
state\cite{goodearl} and does not come as a consequence like in the complex
case\--  precisely means that
$\tau=\omega_1$.  To extend the
field of investigation, one can consider states that do not preserve the
conjugation -- then the supremum is no longer reached by a positive element--
but this contradicts the spirit of density matrices in quantum mechanics.
More interesting is probably to take into account complexified states,
that is real linear functions with value in $\cc$.  

The reduction of
$\aa_I$ to
$\kk^2$ (Proposition 3) is made possible by the orthogonality of the
projections. When
the two internal pure states are no longer orthogonal, there is no
reason why the relevant picture should remain the two-sheets model.
The same is true for two orthogonal states whose sum of the
projections does not commute with the Dirac operator. In this sense, if
these cases do not support a simple "classical" picture (such as
being the geodesic distance of a discrete Kaluza-Klein manifold), they
reflect a purely noncommutative aspect of space-time.

Note that the result -- before fluctuation -- concerning states
defined by the same pure state of one of the algebras (Theorem
\ref{sansfluct}), as well as the reduction from $\aa_I$ to $\kk^2$,
 do not assume that
$\aa_E$ is Abelian. It is only later, to establish the orthogonality
between the
internal and the external spaces, that $T_E$ is taken as the spectral triple of
a manifold. It would be interesting to clarify the importance, or the
unimportance, of the commutativity regarding Pythagorean theorem.

\subsection*{Acknowledgements.}

The authors are grateful to T. Krajewski who introduced several ideas
in the original computation of the two-sheets model, and to B. Iochum
whose help has been essential to generalise and simplify the proofs.
Thanks to D. Perrot for mathematical advice.
\pagebreak

\end{document}